\documentclass[aps,prl,preprint,groupedaddress]{revtex4-2}

\usepackage{graphicx}

\usepackage{siunitx}
\usepackage{amsmath,amsfonts,amssymb}
\usepackage[english]{babel}

\begin{document}


\title{Einstein beams and the diffractive aspect of gravitationally-lensed light}

\author{V Rodr\'{i}guez-Fajardo, T~P~Nguyen, K S~Hocek, J~M~Freedman and E~J~Galvez}%

\affiliation{Department of Physics and Astronomy, Colgate University, 13 Oak Drive, Hamilton, 13346, New York, USA}
\date{\today}\textbf{}
\begin{abstract}
The study of light lensed by cosmic matter has yielded much information about astrophysical questions. 
Observations are explained using geometrical optics following a ray-based description of light. 
After deflection the lensed light interferes, but  observing this diffractive aspect of gravitational lensing has not been possible due to coherency challenges caused by the finite size of the sources or lack of near-perfect alignment. In this article, we report on the observation of these wave effects of gravitational lensing by recreating the lensing conditions in the laboratory via electro-optic deflection of coherent laser light. The lensed light produces a  beam containing regularities, caustics, and chromatic modulations of intensity that depend on the symmetry and structure of the lensing object. We were also able to observe previous and new geometric-optical lensing situations that can be compared to astrophysical observations. This platform could be a useful tool for testing  numerical/analytical simulations, and for performing analog simulations of lensing situations when they are difficult to obtain otherwise. We found that laboratory lensed beams constitute a new class of beams, with long-range, low expansion, and self-healing properties, opening new possibilities for non-astrophysical applications.
\end{abstract}

\maketitle

\section{Introduction}
Research on gravitational lensing over the past 30 years has had tremendous success in understanding the phenomenon itself, and in using it for extracting information about the cosmos. Arcs and rings produced by gravitational lensing are now a common appearance in deep-field astronomical images. It constitutes a telescope to probe objects further away from the reach of telescopes on or around Earth to investigate, for example, ancient galaxies  \cite{AugerApJ10}. One of the greatest successes of gravitational lensing has been to investigate dark matter \cite{TreuApJ04,MasseyRPP10} and to put constraints on the abundance of massive astrophysical compact halo objects (MACHOs) as an explanation for dark matter  around our galaxy \cite{Alcock_2001}. Other important accomplishments include obtaining constraints on the Hubble  constant \cite{CollettPRL19} and the finding of exoplanets around exotic bodies, among other findings, via microlensing \cite{MaoRAA12}. 

The phenomenon is well understood in terms of ray optics, which leads to a very important aspect of gravitational lensing: the magnification of the light from a far-away object. It leads to the observation of a multitude of images of the same object, producing Einstein rings for the case of near-perfect alignment of the source/lensing-mass/Earth;  arc-pairs for slight misalignments; and other arcs and crosses for more general asymmetric lensing objects, such as galaxies or clusters \cite{PLWbook}. The ray analysis of lensing yields caustics that are  well understood in terms of singular optics  \cite{BlandfordARAA92,PLWbook}. 

The diffractive aspect of gravitational lensing, the interference of light waves after deflection, has been discussed since the early days of this field \cite{RefsdalMNRAS64}.  
Gravitational deflections are achromatic \cite{PLWbook}, but the full light field in the observation plane should involve diffraction patterns due to the wave aspect of light. Lensed light is strongly focused forward over a small solid angle.  The earliest inquiry into this question found that monochromatic light produced by a symmetric  lensing object should have the approximate form of a Bessel function \cite{Bliokh1975}, or more precisely expressed in terms of confluent hypergeometric functions \cite{HerltIJTP76,DeguchiApJ86,DeguchiPRD86}. Beam patterns of asymmetric lenses should yield astroid caustics  \cite{BlandfordARAA92} bearing diffractive decorations \cite{BerryJO21}.

For most strong lensing situations with deflections of the order of arc-seconds, the misalignment of  the source, lens, and Earth creates path differences much larger than the coherence length of the light, negating the possibility of wave interference. The finite spatial extent of the sources have the effect of averaging out the fringes even for near-perfect alignment \cite{PetersonApJ91,NakamuraPTPS99}. 

Given the challenges that are faced in observing the diffractive effects of lensing of light waves,  an alternative method to reproduce them via laboratory methods would enhance our understanding of the phenomenon and help with further analysis and discoveries.  Previous attempts at simulating gravitational lensing in the laboratory include the use of logarithmic-shaped axicons \cite{IckeAJP80,HigbieAJP81}, gradient-index media \cite{PhilbinSc08}, metamaterials \cite{GenovNatPhys09}, transformation optics \cite{ChenOE10} and water surface tension \cite{BarceloNatPhys19}. Most of these attempts suffer from being limited by a rigid optical element.

In this work we use computerized holography with an electro-optical device to deflect the light of a laser beam according to the predictions of gravitational theory. The resulting beams are a novel class of beams that adhere to the gravitational lensing conditions. As will be shown below, these beams change radically in shape depending on the structure of the lens, and so to distinguish them from other types of beams we call them ``Einstein beams.'' They enable us to fully simulate gravitational lensing and study them while varying all possible parameters of the problem, with the key advantage that they can be programmed to produce any type of lensing situation.

This paper is organized as follows. In Sec.~2 we give a description of the method to produce gravitational-like deflections of light beams. The results are then presented in two sections, beginning with our observations of the diffractive features of lensed beams in Sec.~3, and continuing by showing that our method also reproduces the geometric-optic astrophysical observations of previous lensing situations plus new ones in Sec.~4. We present experimental details in Sec.~5 and finish with concluding remarks in Sec.~6.

\section{Gravitational lensing in the laboratory}
\begin{figure}
\centering
\includegraphics[width=100mm]{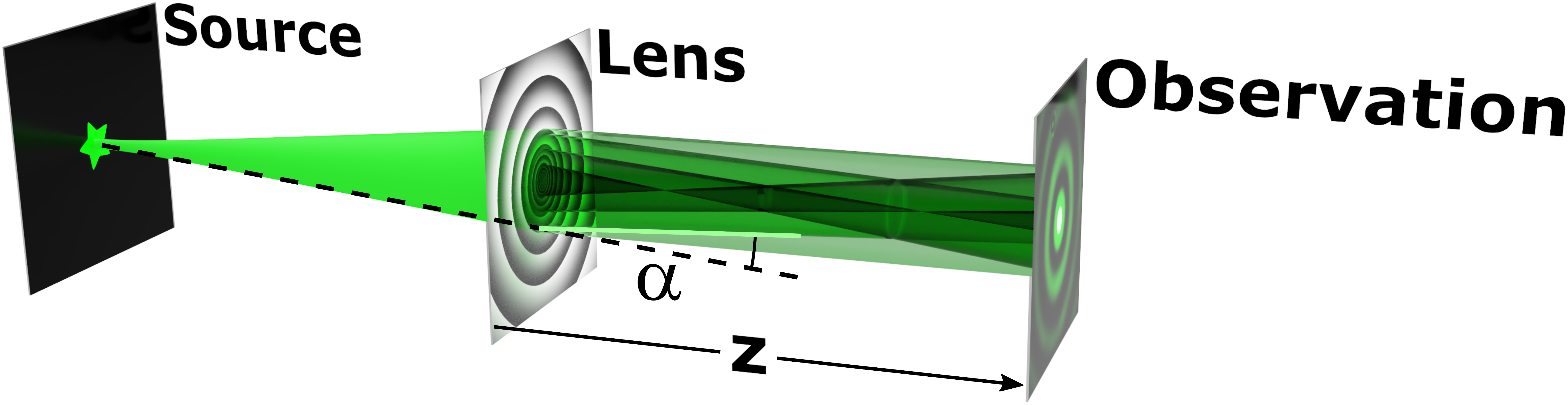}
\caption{Conceptual sketch of the physical system we investigate. Light coming from a distant source is deflected by the gravitational field of a lensing mass through an angle $\alpha$, that depends both on the Schwarzchild radius of and the distance to the deflecting mass. This can be translated to a phase change following a logarithmic dependence on the radial direction. Observation is carried out at a distance z away from the lensing plane.}
\label{fig:concept}
\end{figure}

In gravitational lensing, light deflection occurs over a relatively short range of distances compared to the distances of the source and Earth to the lens \cite{PLWbook}. Thus, it is fair to approximate the deflection as occurring instantaneously on a plane (Fig.~\ref{fig:concept}). The deflection angle $\alpha$ depends inversely on the impact parameter $r$, the distance from the lensing mass to the ray of light in the deflection plane \cite{Schneider}, 
\begin{equation}
\alpha=\frac{2r_S}{r},
\label{eq:alpha}
\end{equation}
where $r_S=2GM/c^2$ is the Schwarzschild radius, with $G$ being the gravitational constant, $M$ the mass of the deflecting object and $c$ the speed of light. Such a deflection was first predicted by Einstein  in 1915 based on general relativity \cite{Einstein15,Einstein16}, and measured by  Eddington and Dyson in 1919 \cite{Dyson20} in a historic episode of scientific discovery.  

A spatial light modulator (SLM), a pixelated computer-programmed liquid-crystal-based phase shifter, can produce the same type of deflections in the laboratory. Observing them along the span of an optical table requires the use of larger deflection angles, of the order of arc minutes. We found that it allowed the investigation of all aspects of the phenomenon. The SLM deflects the light by imparting phase shifts on an incoming light beam. Gravitational-like deflection following Eq.~\ref{eq:alpha} can be imparted onto the light by encoding a position-dependent phase \cite{BerryJO21,GalSPIE}
\begin{equation}
\phi_{\rm SLM}=-2kr_S\ln\left(\frac{r}{r_0}\right),
\label{eq:phislm}
\end{equation}
where $r$ now is the radial coordinate on the SLM, $k$ is the wave number of the light and $r_0$ is a reference radius where $\phi_{\rm SLM}=0$. 
We simulated a number of lensing situations described below.  We programmed values of $r_S$ in the range $0.1-\SI[]{10}{\micro\meter}$, which in astrophysical situations would correspond to lensing masses from $4\times10^{-11}M_{\odot}$ to $3\times 10^{-9}M_{\odot}$, where $M_{\odot}$ is the mass of the Sun. A misaligned situation was simulated by changing the location of the central mass programmed onto the SLM by horizontal and vertical amounts $\delta x$ and $\delta y$.  Asymmetrical deflecting masses, such as  elliptically-shaped galaxies or clusters, were simulated by adding a parameter $e$ that is related to the ellipticity of the lens. Incorporating these features entailed modifying the encoding of Eq.~\ref{eq:phislm} by
\begin{equation}
r\rightarrow r\sqrt{\cos^2\phi+e\sin^2\phi}.
\label{eq:ellip}
\end{equation}  
Lensing by binary mass systems was studied as a function of their mass and separation  by superimposing displaced phase encodings. The possibility that certain astrophysical objects, such as  Kerr black holes, give angular momentum to the light, has been proposed \cite{HarwitAJ2003,TamburiniNP11}, and measured \cite{TamburiniMNRAS20}. Adding an azimuthal phase to the encoding of the form $\ell\phi$, where $\ell$ is the topological charge (an integer) recreated situations where the lensing objects imparted orbital angular momentum of $\ell\hbar$ per photon \cite{Allen92}. 

\begin{figure}
\centering
\includegraphics[width=100mm]{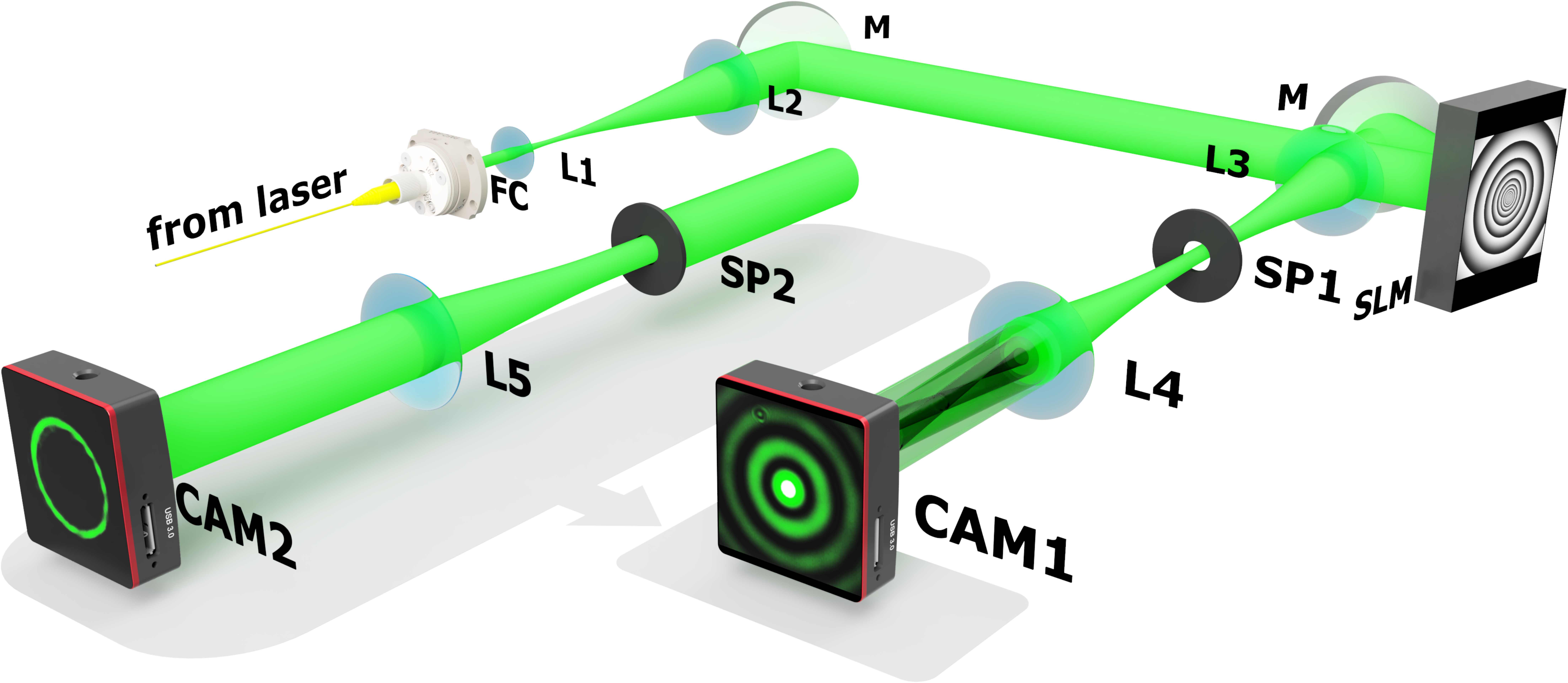}
\caption{Schematic of the laboratory implementation. The input laser beam was optionally filtered by a single-mode fiber connected to a fiber coupler (FC). The (point) source was adjusted by lenses L1 and L2. The light beam was then steered by mirrors (M) to a spatial light modulator (SLM) with a phase encoding that affected the light deflections. The beam was imaged by placing the camera (CAM1) on the light path after the 4f relay system of lenses L3 and L4. Astrophysical observations were mimicked by capturing light that was transmitted by a small aperture (SP2) in the beam path, and far-field imaged by a lens (L5) and camera (CAM2).
}
\label{fig:setup}
\end{figure}

Figure~\ref{fig:setup} shows a schematic of the optical setup. We used coherent light beams in the TEM$_{00}$ mode from one of several laser sources and wavelengths: gas-based He-Ne (\SI[]{633}{\nano\meter}), He-Cd (\SI[]{442}{\nano\meter}) and Ar-ion (458, 477, 488, 496, 501, \SI[]{514}{\nano\meter}); diode-pumped solid-state lasers (532 and \SI[]{589}{\nano\meter}), and  diode lasers at 405, 670 and \SI[]{694}{\nano\meter}.  A system of 2 achromatic lenses (L1 and L2, with focal lengths $f_1=\SI[]{50}{\milli\meter}$ and $f_2=\SI[]{500}{\milli\meter}$, respectively) was used to mimic the point source of light shown, adjusted to produce curved or planar wavefronts reaching the lensing object. For most of the data presented in this article, we used the simpler setting of a planar wavefront incident onto it, which is the case when the source-lens distance is much greater than the lens-observer distance. This is the case shown in Fig.~\ref{fig:setup}, where an optical fiber was connected to a collimator (FC) producing a beam of small size that is later expanded and recollimated by lenses L1 and L2. The beam was then directed towards a liquid crystal spatial light modulator (SLM: Hamamatsu model LCOS, with $792\times 600$ pixels with \SI[]{20}{\micro\meter} pixel size) where a suitable digital hologram was displayed. These were created using MATLAB as images with the desired phase profile (modulo $2\pi$)  encoded as grayscale (adjusted for the wavelength in use). The image seen on the SLM shows an example of such a phase encoding for the case of a symmetric lens. To this phase, we added a phase-grating encoding to produce the lensed beam on the first-order diffraction, plus aberration corrections to counteract any SLM imperfections (not shown). A 4f imaging lens system (L3 and L4, both achromatic with focal length $f_{3,4}=\SI[]{500}{\milli\meter}$, respectively) relayed the deflected beam further away from the SLM for better diagnosis. The entire ``Einstein'' beam, was imaged by placing a camera (Thorlabs DCC1645C with \SI[]{3.6}{\micro\meter} pixel size) along the beam path.  We mimicked the astrophysical observations by limiting the light to pass through a small pinhole of about 0.5-\SI[]{1}{\milli\meter} in diameter. A lens (L5 with focal length \SI[]{100}{\milli\meter}) and a camera (Thorlabs DCC1545M with \SI[]{5.2}{\micro\meter} pixel size) separated by the focal length of the lens provided the scale on the camera plane. Image analysis and fully automated data acquisition were made using custom-made MATLAB scripts.

\section{Diffractive features}
\begin{figure}
\centering
\includegraphics[width=90mm]{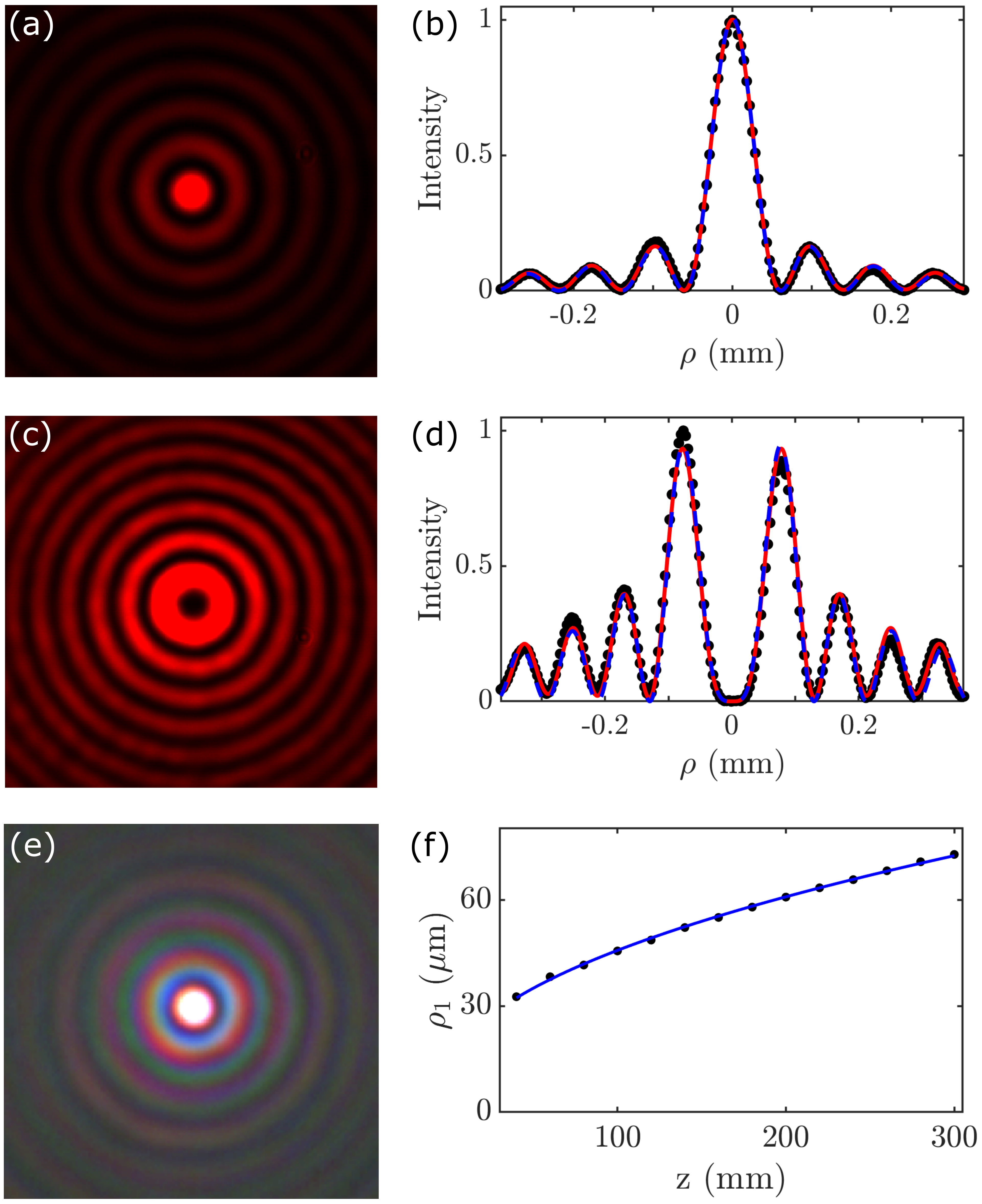}
\caption{Images of Einstein beams with (a) $\ell=0$, and (c) $\ell=2$. They are well explained by confluent hypergeometric functions and Bessel modes of order $\ell$, as shown by the fit for $\ell=0$ in (b) and $\ell=2$ in (d). (e) Image montage of the beam taken at 12 wavelengths (405, 442, 458, 477, 496, 488, 514, 532, 589, 633, 670  and \SI[]{694}{\nano\meter}) with intensity magnified to better appreciate the chromatic character. (f) The radius of the first minimum for $\ell=0$ as a function of the propagation distance.}
\label{fig:eb}
\end{figure}

Placing a camera in the observation plane of the lensed light revealed the wave aspect of gravitational lensing. Figure~\ref{fig:eb}(a) shows the pattern for a symmetric lensing object taken with a monochromatic beam (of wavelength \SI[]{633}{\nano\meter}). 

The measured light beam reveals a pattern with the striking regularity of a Bessel function, confirming previous treatments of the problem \cite{Bliokh1975,HerltIJTP76}. While the shape remains constant, the beam pattern expands non-monotonically with propagation. We obtained the exact shape of the pattern  analytically by propagating a field of the form $U_0\propto\exp[i2r_S\ln(r/r_0)+i\ell\varphi]$ using the Fresnel diffraction integral \cite{goodman}. 
The result is a light mode represented by \cite{RodPSPIE23}
\begin{multline}\label{eq:chf}
U(\rho,\varphi;z) \propto \exp\left(i\left[kz-\ell\left(\varphi+\frac{\pi}{2}\right)+k\frac{\rho^2}{2z}\right]\right) \rho^\ell \\
    \mbox{}_1{\rm F}_1\left(\frac{\ell+2}{2}-ikr_S;\ell+1;-\frac{ik\rho^2}{2z}\right)
    \left(\frac{1}{z}\right)^{\frac{\ell}{2}-2+ikr_S }\,
\end{multline}
where $\rho$ and $\varphi$ are respectively  the radial and azimuthal coordinates in the transverse plane at position $z$ along the propagation direction, and $\mbox{}_1F_1(a,b;x)$ is a confluent hypergeometric function.  
In the asymptotic approximation, where $k\gg1/r_S$,
 the field converges to 
\begin{equation}
U\propto {\rm J}_\ell(k\sqrt{2r_S/z}\;\rho),
\label{eq:ubess}
\end{equation}
which is a Bessel function of order $\ell$.  

The Einstein beam on its own has interesting properties. Analysis of the beams gave good fits of the intensity profile both to the exact and approximated models, as shown in Fig.~\ref{fig:eb}(b), which displays a cut of the \SI[]{633}{\nano\meter} beam along the center plotted with the corresponding fits. Such a pattern corresponds to a Fresnel number $m=2r_S/\lambda=9.5$. Our system allowed values of $m$ within the range 0.2 to 30. When adding orbital angular momentum to the beam, which can be imparted by a Kerr black hole \cite{HarwitAJ2003}, the field  vanishes at the phase discontinuity. Analytically, this is ensured by the term $\rho^\ell$ in the exact solution of Eq.~\ref{eq:chf} and it is consistent with the asymptotic approximation since Bessel functions with $\ell>0$ naturally vanish at $\rho=0$. Fig.~\ref{fig:eb}(c) and (d) show the measured beam and corresponding fit for $\ell=2$, respectively, which is also fit successfully by both theoretical approaches (exact and approximated).

The size of the radius of the first minimum $\rho_{1-{\rm min}}$ is of the order of tens of micrometers. This can also be calculated from Eq.~\ref{eq:ubess}, by setting the argument $k\alpha_E\rho_{1-{\rm min}}=2.405$, where
\begin{equation}
\alpha_E=\sqrt{\frac{2r_S}{z}},
\label{eq:alphae}
\end{equation}
is the angular radius of the Einstein ring when the object-lens distance is much larger than the lens-observer distance ($z$) \cite{GalSPIE}. As mentioned earlier, we programmed our lens for deflections of the order of arc minutes. Astrophysical imaging observations of  lensing in the visible are typically in the order of arc seconds. This puts $\rho_{1-{\rm min}}$ for astrophysical observations in the millimeter range. Thus, observations of lensing with telescopes, such as the Hubble Space Telescope (HST), average through the Einstein-beam fringes, if at all observable. For a fixed lensing parameter $\alpha_E$, the radii of the rings increase linearly with the wavelength. A composite image of 12 wavelengths across the visible is shown in Fig.~\ref{fig:eb}(d). Given that the radii of the rings increase at different rates for different wavelengths, the overall pattern is very colorful due to the overlap of minima at some wavelengths with maxima at other wavelengths.  It constitutes an illustration of the dispersion caused by the  diffraction of gravitationally lensed light. 

Einstein beams can also be considered a new class of light beams. They are an intermediate case between non-diffracting beams, such as Bessel beams \cite{DurninPRL87}, which show no expansion but are short-ranged; and Gaussian beams, which are long-ranged but expand asymptotically linearly with $z$. We confirmed the prediction of Eq.~\ref{eq:chf} of a dependence that increases with $\sqrt{z}$. In Fig~\ref{fig:eb}(f) we display a graph of the radius of the first minimum of the fitted pattern along with a fit proportional to $\sqrt{z}$, where we see the beam expand from \SI[]{36}{\micro\meter} to \SI[]{67}{\micro\meter} in \SI[]{0.25}{\meter}.  Einstein beams can be long-ranged.  The range $z_{\rm max}$ of the Einstein beam can be calculated roughly by the lowest deflection angle at the rim of the SLM ($r_{\rm SLM}=\SI[]{6}{\milli\meter}$) for incoming parallel rays, yielding $z_{\rm max}=r_{\rm SLM}^2/(2r_S)$. For example, using $r_S=\SI[]{0.5}{\micro\meter}$ we get $z_{\rm max}=\SI[]{36}{\meter}$. However, our uncertainty in the expansion of the input beam can make the range vary significantly, a slight expansion of the incoming beam can significantly vary the range. We have observed a pattern featuring at least 3 rings, with $\rho_{1-{\rm min}}=\SI[]{7}{\milli\meter}$ at a distance of about \SI[]{100}{\meter} along the hallways adjacent to our lab. A Gaussian beam starting with a waist of \SI[]{36}{\micro\meter} has a much larger beam radius of about \SI[]{530}{\milli\meter} at that distance.

There is potentially a new associated phenomenon to be found in astrophysical observations: self-healing of the light beam from the shadow of obstacles. Einstein beams and Bessel beams are made by the intersection of conical rays. If we imagine a plane that contains the rays and the beam axis, in the case of Bessel beams, the rays coming from above the axis are all parallel, and similarly rays coming from below. In Einstein beams those sets of rays are not parallel. Instead, they have slowly varying slopes, as shown schematically in Fig.~\ref{fig:concept}. At different distances along the propagation direction the Bessel-beam mode is made approximately of a finite set of rays, but beyond a self-healing distance, the pattern is mostly made of a different set of rays. Thus, an obstacle in the path of the beam interrupts a set of rays creating a shadow, but further downstream the shadow disappears, and the beam  self-heals \cite{BouchalOC98}. The same is true for Einstein beams. This is not only a curious phenomenon, but the actual case for gravitational lensing, where objects in space along the beam's path act as obstacles. In an aligned system for a light source sufficiently far away that the light rays arriving at the lensing plane can be assumed to be parallel to the optical axis, using geometrical considerations results in a self-healing distance $z_{SH}=x/(2\beta)$, where $x$ is the physical size of the obstacle and $\beta$ is the angle that the rays form with the axis \cite{BouchalOC98} (which is constant for Bessel beams). In Einstein beams $\beta=\sqrt{2r_S/z}$, where $z$ is the distance from the lens to the self-healing point, thus $z_{SH}$ increases with object size and with the separation between the obstacle and the lens. 
We verified the self-healing effect but defer details for a separate publication. The self-healing images are similar to what one expects of Bessel beams. 

\begin{figure*}
\centering
\includegraphics[width=\textwidth]{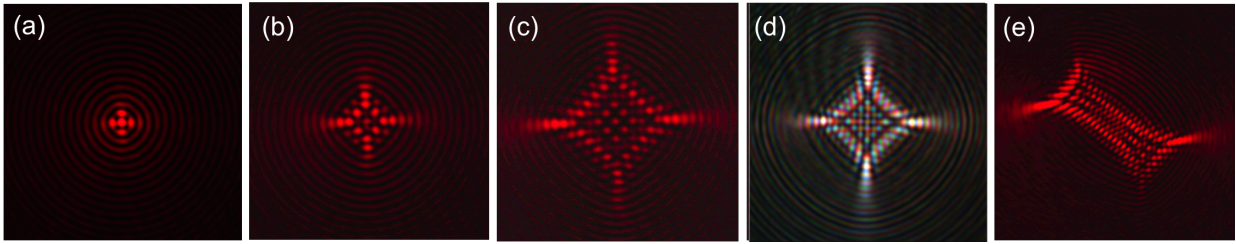}
\caption{Mosaic of Einstein beams bearing asymmetry. Elliptical lenses with $2kr_S=20$:  for (a) $e=1.1$, (b) $e=1.3$ and (c) $e=2$. (d) Image montage of an elliptical lens with $e=1.5$ and $r_S=\SI[]{5.5e-6}{\meter}$ (488, 514, 589, $r_S=\SI[]{964}{\nano\meter}$). (e) Elliptical lens of (c) plus orbital angular momentum with $\ell=15$.}
\label{fig:assym}
\end{figure*}

Asymmetric lensing objects have a significant effect on Einstein beams. The smallest asymmetry transforms the symmetric Bessel-like pattern into astroid patterns made of four cusps, and because of the monochromatic source that we use, it features  lattice interference patterns. The pattern that we see with symmetric lensing has an axial caustic, and the Bessel rings are symmetric interference fringes. The asymmetry of the lens produces crossings of gravitationally deflected light that results in 2-dimensional interference lattice points, or decorations \cite{BerryJO21}.  In Fig.~\ref{fig:assym}(a) we see the smallest effect of the elliptical perturbation:  the coalescence of the first ring  into 4 interference maxima aligned with the axes of the asymmetry. We observed the same type of pattern with elliptical lenses and binary lenses of low separation.

As the ellipticity of the elliptical lens is increased, and similarly, the separation of binaries, the two types of patterns undergo a transformation. The next outer complete ring from the previous set of parameters coalesces into a set of lattice points forming a diamond, while the internal set of interferences fades, as shown in Figs.~\ref{fig:assym} (b) and (c). It is also seen that the ends of the diamond are cusps due to a caustic fold \cite{BerryJO21,Nye}. The coalescing of rings into lattices proceeds as we increase the ellipticity further. The sequence of patterns proceeds at different rates for different wavelengths such that the patterns do not overlap. In Fig.~\ref{fig:assym}(d) we show a pattern for a select number of wavelengths.  Such chromatic effects  of lensing have never been observed. 
The addition of orbital angular momentum to the elliptical lens distorts the symmetric diamond into a displaced rectangular pattern, as shown in Fig.~\ref{fig:assym}(e) for $\ell=-15$. The pattern mirror-flips about a vertical axis for $\ell=+15$.


\section{Geometrical-optics Features}
\begin{figure}
\centering
\includegraphics[width=0.7\textwidth]{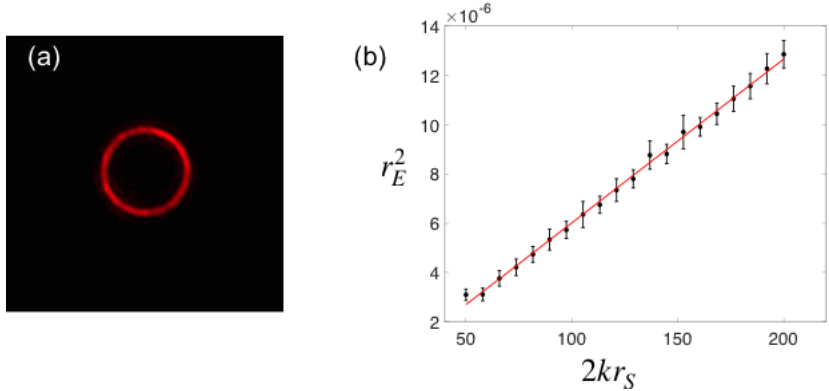}
\caption{(a) Image of the laboratory recreation of an Einstein ring;  (b) Linearized graph of the relation of the Einstein-ring diameter with the Schwarzchild radius $r_S$, which is effectively the mass of the lensing object.}
\label{fig:eringvrs}
\end{figure}

The most notorious but less common situation in gravitational lensing involves the fortuitous alignment of object, lens, and observer, which yields a symmetric Einstein ring. Fortunatley, the cosmos is so vast that Einstein rings still abound \cite{MiraldaEscudeMNRAS92}, and have been observed since 1988 in the radio \cite{HewittNature88} and in 1992 in the visible \cite{WarrenAA92}, with numerous others found since, for example,  via the Sloan Sky survey \cite{Bolton08}. The angular  Einstein radius at a distance $z$ from the lens is given by Eq.~\ref{eq:alphae}.
To mimic astrophysical observations, which sample a small region of space, and integrate it over the telescope aperture, we place a $\sim\SI[]{1}{\milli\meter}$ aperture and an additional lens in the path of the beam (See Fig.~\ref{fig:setup}). Figure~\ref{fig:eringvrs}(a) shows an example of a laboratory Einstein ring. For this image we had $\alpha_E=4.9\times10^{-3}\;{\rm rad}=17$ arc min for $k\simeq\SI[]{e7}{\per\meter}$ and $r_S=\SI[]{3e-6}{\meter}$. It produced a \SI[]{0.5}{\milli\meter} radius ring on the  active area of a CMOS digital camera. The predictions of Eq.~\ref{eq:alphae} were confirmed  by measuring their radius and graphing it as a function of the  lensing mass, as shown in  Fig.~\ref{fig:eringvrs}(b) in a linearized plot. A linear fit to the data confirmed the relation. In a previous study, we also confirmed the non-linear dependence of the Einstein radius with $z$ \cite{GalSPIE}.

\begin{figure*}[tb]
\centering
\includegraphics[width=0.8\textwidth]{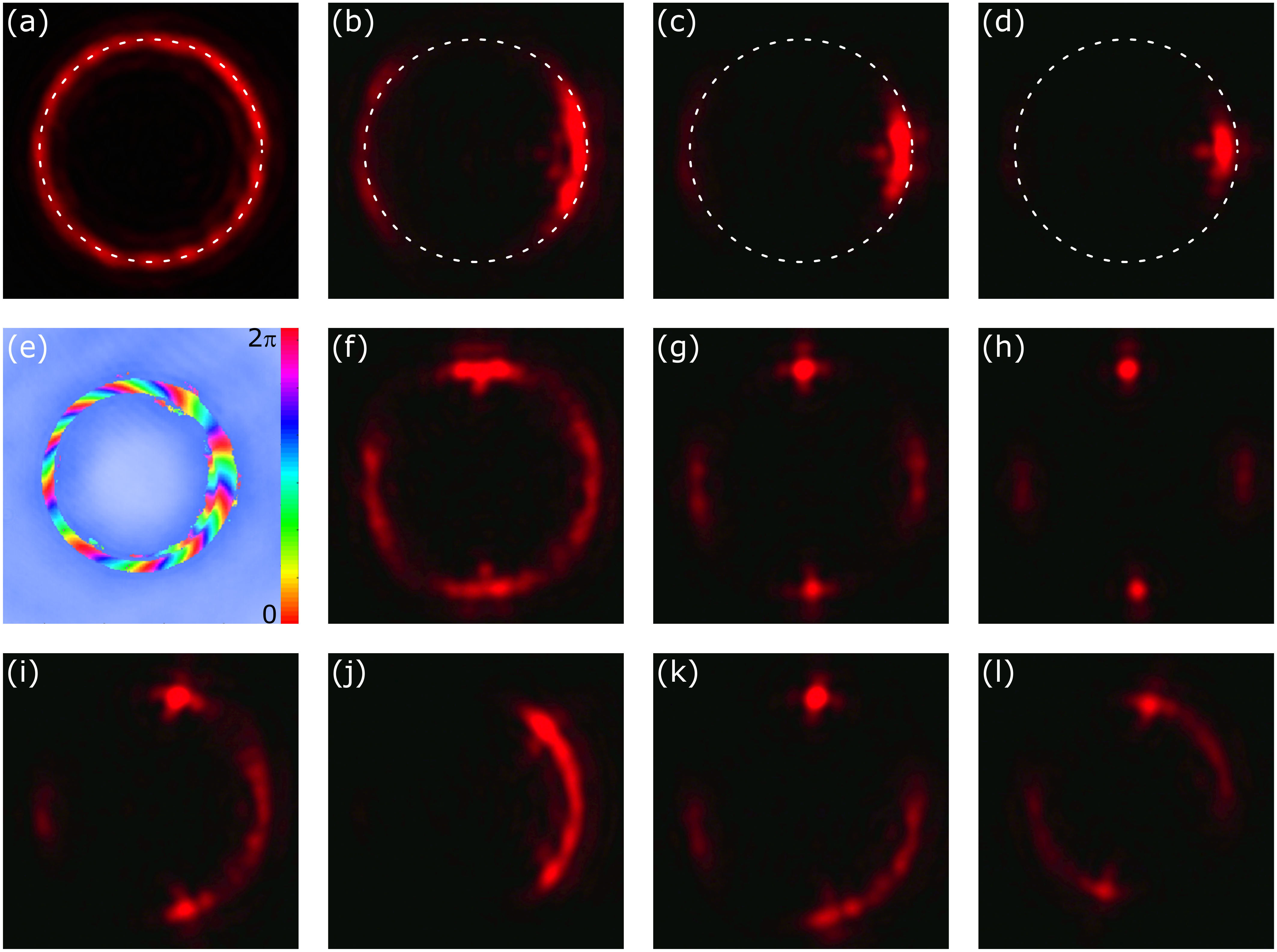} 
\caption{Mosaic of observations of gravitational lensing as a function of laboratory parameters. For symmetric lensing objects, we show a sample of the measurements as a function of the displacement of the center of the lensing mass programmed in the SLM with values (a) $\delta x=0$, (b) $\delta x=15$, (c) $\delta x=25$ and (d) $\delta x=35$ pixel units of the SLM. Dashed circles mark the position of the Einstein ring. In all images but (a) the intensity was artificially increased beyond saturation to make the fainter features more visible. (e) False color phase map of a ring with orbital angular momentum with $\ell=20$. An asymmetry is added by varying $e$ in Eq.~\ref{eq:ellip} for values (f) $e=1.3$, (g) $e=1.5$ and (h) $e=2$. Addition of a displacement to the asymmetric lens with $e=1.5$: for (i) $\delta x=15$ px, (j) $\delta x=50$ px, and (k) $\delta x=\delta y=10$ px. (l) Addition of orbital angular momentum with $\ell=10$ for $e=1.5$ with no displacement.}
\label{fig:earcs}
\end{figure*}

Displacing the center of the hologram representing the lensing mass reproduced well-known Einstein arcs. These are stationary points in the optical path length from the object to the observer, which for a symmetric lens results in 2 arcs \cite{NandorAJP96}. Such features of lensing were the first evidence of lensing in astrophysical observations \cite{WalshNature79}, and are ubiquitous in deep-sky images showing gravitational lensing, including spectacular sights obtained using the HST and other telescopes. The distinct path lengths that produced the two arcs are of much interest in astrophysics because the light from the separate arcs involve delay times that have been used for determining a more accurate value of the Hubble constant \cite{RefsdalMNRAS64,CollettPRL19}. By being able to vary the misalignment smoothly in our laboratory measurements, we were able to reproduce the smooth change of the ring into arcs, as shown in Fig.~\ref{fig:earcs}(a-d), showing the transition for selected values of the misalignment. A dashed outline of the Einstein ring obtained for no displacement (Fig.~\ref{fig:earcs}(a)) is used to measure the change in position of the arcs as a function of the displacement. We observe that as the displacement increases, both arcs change in size, position  and in relative intensity.

We tested the orbital angular momentum of an Einstein ring, by adding to the lens a topological charge $\ell$, as described earlier, and setting it up in an interferometer where the Einstein ring bearing orbital angular momentum interferes with a plane wave. By varying the relative phase we were able to produce  interferograms and confirm that the  topological charge imparted by the lens is encoded onto the ring, as shown in Fig.~\ref{fig:earcs}(e), in a false-color image of the phase in a ring determined for the case where $\ell=20$. 

When we added elliptical asymmetry to the lensing mass, we observed the Einstein cross, which is also explained in terms of stationary points aligned along the axes of the elliptical lens. In Fig.~\ref{fig:earcs} we show images for $e=1.3$ (f), $e=1.5$ (g) and $e=2$ (h). where $e$ is as defined in Eq.~\ref{eq:ellip}. In that sequence, we can appreciate the metamorphosis of the ring into arcs. We observe that they transform by increasing the intensity of one opposite pair of arcs relative to the other.  The horizontal and vertical axes flip for $e<1$.

If we add a displacement along the horizontal direction, which is one of the most common misalignment situations in astrophysical observations \cite{Bolton08}, we observe the merging of 3 stationary points into a continuous long arc plus a short arc or point opposite to it, as seen in Fig~\ref{fig:earcs}(i) for $\delta x=15 $ pixels (px). The single arc decreases in size for increasing displacements, such as $\delta x=50 $~px seen in (j). A displacement along 2 directions, such as $\delta x=\delta y=10 $~px in (k), shows yet a new situation: 2 arcs merge to form a 3-arc pattern \cite{BlandfordARAA92}. Observing such dynamic evolution has not yet been seen astrophysically, but is easily done with our system.

For binary lenses, we see similar effects to those of elliptical lenses, thus we do not show them here for the sake of brevity. In this case, however, the Einstein ring turns into an ellipse as the separation between the binaries is increased from zero, and coalesces into a cross as the separation is increased further.  
Our parameter set allowed us only a maximum distance between binary masses of about $10^4r_S$, more resembling the situation of contact binaries or black-hole/neutron-star systems.
Adding orbital angular momentum results in adjacent pairs of arcs of the Einstein cross to merge into 2 wide pairs, as shown in Fig.~\ref{fig:earcs}(l) for a topological charge $\ell=10$. This has not been observed or recognized, as far as we know. Reversing the sign of the charge alternates the pairing of arcs. 

If we decrease the mass of the lensing object so that the ring is no longer seen due to its radius being of the order of the resolution of the optical system, we get into the microlensing-like regime. We could simulate weak lenses down to $r_S=\SI[]{0.025}{\micro\meter}$, being the SLM's bit-depth the limiting factor. We mimicked microlensing events as we scanned the displacement of the lens $\delta x$ from -100 px to +100 px, and observe an increase in the measured intensity, peaking at $\delta x=0$, and decreasing as $\delta x$ was increased further. 
The arcs are more resilient to obstacles in self-healing, but show deformations and caustics due to asymmetries in the obstacle. This could also be used to study the effect of complex caustics in microlensing \cite{Schneider}.

\section{Discussion and Conclusions}
In summary, we have implemented a method to observe all of the gravitational lensing effects in the laboratory using spatial light modulation. Astrophysical observations span a wide range of scales, from wavelengths at the short end of the visible spectrum to the radio. Our laboratory recreations tested a slice of this parameter space. With simulated gravitational deflections of the order of arc minutes and coherent visible light from a laser, we could image the beam onto a standard digital camera. This revealed patterns with the striking regularity of Bessel-like beams for symmetric lenses. It allowed us to observe the predicted caustics of gravitational lensing produced by asymmetric lenses, which show lattice interference patterns that are rich and colorful when the source of composed of multiple wavelengths. These types of details are not possible in astrophysical observations because of the stringent conditions that are required, such as a point-like source, perfect alignment, and the right combination of telescope aperture and wavelength of the radiation. There is still hope to observe wave effects with candidate point sources, such as pulsars, gamma-ray bursts\cite{JowMNAS20}  and fast radio bursts  \cite{KatzMNRAS20}. Source size also approximates the point source for near solar lenses \cite{HeylMNRAS11}. The alignment situations where the path difference of the light around the lens is of the order of the wavelength of the light can be achieved with  small masses, such as primordial black holes \cite{Katz_JCAP2018,SugiyamaMNRAS20}. The laboratory capability presented here allows us to explore the gravitational diffraction patterns  that have not yet been observed. Gravitational lensing deflections are independent of the wavelength of the light, but other features of astrophysical lenses, such as plasma surrounding them \cite{TsupkoMNRAS20} or wavelength-dependent masses \cite{WambsganssApJ91} may be simulated with our system. So far these chromatic effects, or more generally diffractive effects, have not been found in astrophysical observations.

While we have only experimented with point light sources, the laboratory capability also opens new possibilities for observing rings and arcs by analog simulations of more complex situations. For instance, additional modulator(s) can be added to simulate extended sources of arbitrary shape. Our images of rings and arcs integrate over about 12 Einstein-beam rings, which is comparable to HST. These types of experiments could allow a way to test current lensing theories and software \cite{BlandfordARAA92,MeneghettiMNRAS17}. Laboratory methods could also be used to study  weak lensing due to dark matter or exoplanets by analog simulations. The question of self-healing is an intriguing one. Could some observations involve self-healing due to an obstruction in the path of the light, and what could they reveal about the obstruction? 

Einstein-type beams can also be used in non-astrophysical applications, such as in light-sheet  microscopy \cite{HuiskenSci04}, remote sensing, and communications due to their low expansion, self-healing, potentially deep penetration, and other properties already considered for logarithmic axicons \cite{GolubOL10,KotlyarJOSAA11}. The SLM can be used in the investigation of a class of beams where the deflection angle is more generally proportional to $r^{-n}$ ($n\in\mathbb{R}_{>0}\ne1$), which may reveal new interesting optical-beam properties such as those described here.

\begin{acknowledgments}
The authors thank M. Alonso, E. Brasselett, S. Francke-Arnold, W. Miller, and J. Pilawa for their help and useful suggestions. This work was funded by NSF grant PHY-2011937.
\end{acknowledgments}

\section*{References}
\bibliography{gravlensing22}


\end{document}